\documentclass{elsart}
\usepackage{graphicx,amsmath,cite,textcomp}
\usepackage{array,amssymb,pifont}
\usepackage{layout}
\usepackage{multirow}
\usepackage{morefloats}
\newcommand{\mat}[2][rrrrrrrrrrrrrrrrrrrrrrrrrrrrrrr]{\left(
    \begin{array}{#1}
    #2\\
    \end{array}
    \right)}
\begin{document}
\begin{frontmatter}
\title{Reconstruction of nuclear charged fragment trajectories from a large gap sweeper magnet}
  \author[NSCL,PA,CC]{N.~Frank\corauthref{cor}},
  \author[NSCL]{A.~Schiller},
  \author[NSCL]{D.~Bazin},
  \author[NSCL,PA]{W.A.~Peters}, and
  \author[NSCL,PA]{M.~Thoennessen}
  \address[NSCL]{National Superconducting Cyclotron
  Laboratory, Michigan State University, East Lansing, MI 48824, USA}
  \address[PA]{Dept. of Physics \& Astronomy, Michigan State University, East Lansing, MI 48824, USA}
  \address[CC]{Dept. of Physics, Concordia College, Moorhead, MN 56562, USA}
  \corauth[cor]{Corresponding author. Tel.: +1-218-299-3236; fax: +1-218-299-4308.\\
  \textit{E-mail address:} nfrank@cord.edu}
\begin{abstract}
A new method to reconstruct charged fragment four-momentum vectors from measured trajectories behind an open, large
gap, magnetic dispersion element (a sweeper magnet) has been developed. In addition to the position and angle behind
the magnet it includes the position measurement in the dispersive direction at the target. The method improves the
energy and angle resolution of the reconstruction significantly for experiments with fast rare isotopes, where the beam
size at the target position is large.
\end{abstract}
  \begin{keyword}
    spectrometer
    \PACS 25.60.-t, 29.30.-h, 29.30.Ep
  \end{keyword}
\end{frontmatter}

\section{Introduction}
The frontier of nuclear physics is moving to more and more neutron rich systems extending out to the limit of existence
(dripline). Only a few if any bound excited states exist in nuclei close to the neutron dripline and beyond, none at
all. The unbound state(s) of nuclei near the neutron dripline are energetically above one- or even two-neutron
separation energies. Thus these nuclei break-up by neutron emission, which makes gamma spectroscopy an ineffective tool
for the study of these nuclei. Neutron unbound states have to be reconstructed by neutron decay spectroscopy
\cite{naka_ref,pain_ref,adrich_ref,thoen_ref}.

Often, the neutron-rich nuclei of interest are produced via one or more nucleon stripping reactions from rare-isotope
beams produced in fast fragmentation reactions. The lifetime of these systems is extremely short ($\sim$10$^{-21}$s) so
they decay within the target immediately after production. Nuclei produced by the fast fragmentation technique have
high energy ($\sim$100~MeV/A) and intensity (10$^{4}$/s) and, beams of such nuclei in general have a large spread in
momentum, size, and angle. Due to the high energy of the beams, the charged fragments as well as the neutrons are
extremely forward focused and have to be detected around zero degrees. The decay energy from neutron unbound states can
be reconstructed from a measurement of the four-momenta of the decay products (charged fragment and neutron).

One method of detecting both the charged fragments and neutrons in coincidence is to separate the charged fragments
from the neutrons using a dipole magnet. In order to maximize the angular acceptance of the neutrons at zero degrees
the dipole (or sweeper) magnet has to have a large gap, an opening at zero degrees (``C"-magnet), and has to be located
as close to the target as possible. These stringent requirements do not allow the use of a full magnetic spectrometer
which is typically used to determine the energy and angles of emitted particles in nuclear structure experiments
\cite{dragon_triumf_ref,grandraiden_riken_ref,s800_ref1,msp144_dubna_ref,bigbite_kvi_ref,k600_ithemba_ref,q3d_hmi_ref}.
For these spectrometers ion-optical codes have been developed to extract the four-momentum of the fragments at the
target from the measured quantities following the spectrometer. One example for such a code which is extensively used
is \textsc{cosy} \textsc{infinity} \cite{berz_ref,berz_ref2,s800_ref2}. However, as we will discuss below, the standard
application of \textsc{cosy} \textsc{infinity} to the specific problem of track reconstruction for the case of a
simple, open, large-gap magnet and a large beam-spot size, which are typical for neutron-decay experiments, does not
yield satisfactory results in terms of resolution.

\section{Method}
\subsection{Fully Inverse Ion-Optical Matrices}
A forward ion-optical matrix relates coordinates ($x$, $\theta_{x}$, $y$, $\theta_{y}$, $\delta$)$^{(T)}$ at the
position of the reaction target in front of the sweeper magnet to coordinates ($x$, $\theta_{x}$, $y$, $\theta_{y}$,
$\Delta L$)$^{(D)}$ at the position of the charged particle detector. In this work, $x$ is the dispersive and $y$ is
the non-dispersive direction of the sweeper magnet. The angular coordinates correspond to ratios of transverse to total
momentum in their respective directions, e.g., $\theta_{x} = p_{x}/p_{0}$. $L$ is the track length of a charged
particle along the central trajectory from the reaction target position to the detector position. Thus $\Delta L$ is
the difference in the track length compared to the length of the central trajectory. The central trajectory is defined
by ($x$, $\theta_{x}$, $y$, $\theta_{y}$, $\delta$)$^{(T)}=0$ and ($x$, $\theta_{x}$, $y$, $\theta_{y}$)$^{(D)}=0$. The
relative energy deviation $\delta$ is given by

\begin{equation}\label{e0}
    \delta = \frac{E - E_{0}}{E_{0}},
\end{equation}

where $E$ is the energy of the incoming particle and $E_{0}$ is the reference energy of a particle following the
central trajectory. This energy is determined by the measured magnetic field of the sweeper magnet.

In first order, coordinates belonging to the dispersive direction and the non-dispersive direction do not depend on
each other because of the magnetic field's symmetry in the non-dispersive direction. Thus, a typical first-order
forward ion-optical matrix will look like the following

\begin{equation}\label{first_order}
\mat{x\\ \theta_{x}\\ y\\ \theta_{y}\\ \Delta L}^{(D)} =
\left(\begin{array}{ccccc}
M_{x x} & M_{x \theta_{x}} & 0 & 0 & M_{x \delta} \\
M_{\theta_{x} x} & M_{\theta_{x} \theta_{x}} & 0 & 0 & M_{\theta_{x}
\delta}\\
0 & 0 & M_{y y} & M_{y \theta_{y}} & 0\\
0 & 0 & M_{\theta_{y} y} & M_{\theta_{y} \theta_{y}} & 0\\
M_{L x} & M_{L \theta_{x}} & 0 & 0 & M_{L \delta}\end{array} \right)
\mat{x\\ \theta_{x}\\ y\\ \theta_{y}\\
\delta}^{(T)},
\end{equation}

where there are two non-zero $2\times2$ sub-matrices $M'$. The sub-matrices mix position and angle in either the
dispersive or non-dispersive direction, respectively. For mixing position and angle in the same direction, $\det M' =
1$ is required in order to preserve the phase-space volume.\footnote{With the given beam energies and intensities,
non-conservative effects from space charges and photon production are estimated to be negligible.} In order to
adequately describe the sweeper magnet's field, the higher-order dependence of output coordinates on combinations of
input coordinates is necessary. For this purpose, the input vector is extended to all possible higher-order
combinations and the ion-optical matrix is modified accordingly. In first order, there are only five linear input
coordinates, hence the ion-optical matrix is a $5\times5$ matrix. In second order, there are 15 new quadratic input
coordinates such as ($(x^{(T)})^{2},x^{(T)} \theta_{x}^{(T)}$); the ion-optical matrix has to be extended by a
$5\times15$ matrix. In third order, there are 35 new third-order input coordinates and the ion-optical matrix has to be
extended by a $5\times35$ matrix and so on. The ion-optical matrices produced and used in this work are up to third
order.

In typical spectrographic applications, the position and angle of particles at the charged particle detector position
are used to determine the energy, angle, and position of those particles at the reaction target. For this reason, the
forward ion-optical matrix has to be inverted. However, looking at Eq. (\ref{first_order}), one notices that a full
inversion is not useful because typically only four out of the five coordinates of the particle are known behind the
spectrograph (the track length is not a measured quantity). This fundamental problem has been solved in two different
ways: (i) if focusing magnetic elements are available, the spectrograph can be run in dispersion-matched mode, in which
essentially the matrix elements $M$ relating the input coordinate $x^{(T)}$ to any of the output coordinates are
negligibly small \cite{s800_ref2}, or (ii) the spectrograph can be run in focused mode, in which the incoming beam is
focused in the dispersive direction at the target, such that one can assume $x^{(T)} = 0$. Either way, Eq.
(\ref{first_order}) can effectively be reduced to

\begin{equation}\label{first_order_redu}
\mat{x\\ \theta_{x}\\ y\\ \theta_{y}}^{(D)} =
\left(\begin{array}{cccc}
M_{x \theta_{x}} & 0 & 0 & M_{x \delta} \\
M_{\theta_{x} \theta_{x}} & 0 & 0 & M_{\theta_{x} \delta}\\
0 & M_{y y} & M_{y \theta_{y}} & 0\\
0 & M_{\theta_{y} y} & M_{\theta_{y} \theta_{y}} & 0\end{array}
\right)
\mat{\theta_{x}\\ y\\ \theta_{y}\\
\delta}^{(T)},
\end{equation}

where the column concerning the target coordinate $x^{(T)}$ and the row concerning the track length variation $\Delta
L$ have been eliminated. After inverting this matrix, the coordinates $(\theta_{x}, y, \theta_{y}, \delta)^{(T)}$ can
be reconstructed from position and angle measurements by the charged particle detectors. The program \textsc{cosy}
provides this kind of inverse ion-optical matrix. In this work, these matrices will be called \emph{fully inverse
ion-optical matrices}.

If it is experimentally not feasible to run in dispersion-matched or focused mode, the above assumption of $x^{(T)} =
0$ is not a valid assumption. The use of the fully inverse ion-optical matrix in the standard \textsc{cosy} method will
not result in sufficient resolution of the reconstructed four-momentum. In these cases it is necessary to measure the
beam position $x^{(T)}$ at the target and include this information in the reconstruction procedure.

\subsection{Partial Inverse Ion-Optical Matrices}
The reaction target coordinates $x^{(T)}$ and $y^{(T)}$ can be measured with tracking detectors located in front of the
target. With these two additional pieces of information, the problem of reconstruction becomes overdetermined. In the
following, the measured target coordinate $x^{(T)}$ is used as an additional input and entered into the reconstruction.

The starting point for our discussion is the first-order ion-optical matrix which relates only the coordinates in the
dispersive direction

\begin{equation}\label{first_order2}
\mat{x\\ \theta_{x}\\ \Delta L}^{(D)} = \left(
\begin{array}{ccc}
M_{x x} & M_{x \theta_{x}} & M_{x \delta} \\
M_{\theta_{x} x} & M_{\theta_{x} \theta_{x}} & M_{\theta_{x}
\delta}\\
M_{L x} & M_{L \theta_{x}} & M_{L \delta} \end{array} \right)
\mat{x\\ \theta_{x}\\ \delta}^{(T)}.
\end{equation}

Here, the target coordinate $x^{(T)}$ is an input to the matrix multiplication on the right-hand side, and the detector
coordinates $x^{(D)}$ and $\theta_{x}^{(D)}$ are outputs on the left-hand side. All three are measured quantities. The
unknown quantities are the target coordinates $\theta_{x}^{(T)}$ and $\delta$ on the right-hand side and the deviation
in track length from the length of the central trajectory $\Delta L$ on the left-hand side. It is now straightforward
to exchange one coordinate on the left-hand side with one coordinate on the right-hand side by using the rules shown in
Table \ref{t_rules}. The choice of the coordinates which one would like to exchange determines the \emph{pivot}, i.e.,
the matrix element which relates the two coordinates.

\begin{table}[p]
  \centering
  \caption{The rules of coordinate exchange for a first order matrix.}\label{t_rules}
  \begin{tabular}{|c|c|}
  \hline
  \emph{pivot} & a $\rightarrow$ 1/a \\\hline
  same row elements as the \emph{pivot} & b $\rightarrow$ -b/a \\\hline
  same column elements as the \emph{pivot} & c $\rightarrow$ c/a \\\hline
  all other elements & d $\rightarrow$ d - cb/a\\
  \hline
  \end{tabular}
\end{table}

The ion-optical matrix

\begin{equation}\label{first_order_exch1}
\mat{\delta^{(T)}\\ \theta_{x}^{(D)}\\ \Delta L^{(D)}} = \left(
\begin{array}{ccc}
-\frac{M_{x x}}{M_{x \delta}} &
-\frac{M_{x \theta_{x}}}{M_{x \delta}} &
\frac{1}{M_{x \delta}} \\
M_{\theta_{x} x}-\frac{M_{x x} M_{\theta_{x} \delta}}{M_{x \delta}} &
M_{\theta_{x} \theta_{x}}-\frac{M_{x \theta_{x}}
M_{\theta_{x} \delta}}{M_{x \delta}} &
\frac{M_{\theta_{x} \delta}}{M_{x \delta}}\\
M_{L x}-\frac{M_{x x} M_{L \delta}}{M_{x \delta}} &
M_{L \theta_{x}}-\frac{M_{x \theta_{x}} M_{L \delta}}{M_{x \delta}}
&
\frac{M_{L \delta}}{M_{x \delta}} \end{array} \right) \mat{x^{(T)}\\
\theta_{x}^{(T)}\\ x^{(D)}}
\end{equation}

is generated by exchanging the reaction target coordinate $\delta$ with the detector coordinate $x^{(D)}$. In a second
step, one can now exchange the reaction target coordinate $\theta_{x}^{(T)}$ and the detector coordinate
$\theta_{x}^{(D)}$ using the same rules, such that the three known quantities are on the right-hand side and the three
unknown quantities are on the left-hand side.

Ion-optical matrices, where coordinates have been exchanged in this fashion will be called \emph{partial inverse
ion-optical matrices} in this work. It is straightforward to generalize this method to the $5\times5$ matrices of first
order. At this point, we would like to make three comments: (i) exchanges of two coordinates are only possible if the
\emph{pivot} does not equal zero, i.e., it is only possible to exchange coordinates in the dispersive plane and in the
non-dispersive plane among themselves. (ii) numerically, the method works better the larger a \emph{pivot} element is
chosen. Ideally, the \emph{pivot} element that is chosen should be greater than one. (iii) when exchanging all five
coordinates along the diagonal, the resulting ion-optical matrix is the mathematical inverse of the original
ion-optical matrix, i.e., in first order, the result of the inversion method is mathematically exact.

The pertinent rules for exchanging coordinates in second or higher order can be derived from the following
considerations. Assume the input, i.e., target coordinates of a forward ion-optical matrix are ($x_{i}$) and the
output, i.e., detector coordinates are ($y_{j}$) with $i, j = 1 ... 5$, then the coefficients of the ion-optical matrix
are $a^{(j)}_{n_{1},n_{2},n_{3},n_{4},n_{5}}$, such that

\begin{center}
\begin{equation}\label{e_higher}
    y_{j=1...5} = \displaystyle\sum_{1 \leq n_{1}+n_{2}+n_{3}+n_{4}+n_{5} \leq N}
    a^{(j=1...5)}_{n_{1},n_{2},n_{3},n_{4},n_{5}} x_{1}^{n_{1}} x_{2}^{n_{2}} x_{3}^{n_{3}} x_{4}^{n_{4}} x_{5}^{n_{5}}
\end{equation}
\end{center}

where $N$ is the order of the ion-optical matrix and $n_{(i=1...5)}\geq0$. Without loss of generality, consider only
the exchange of the target coordinate $x_{5}$ with the detector coordinate $y_{5}$. Then, the matrix elements
$b^{(j)}_{n_{1},n_{2},n_{3},n_{4},n_{5}}$ defined by

\begin{center}
\begin{equation}\label{e_invert1}
y_{j=1...4} = \displaystyle\sum_{1 \leq m_{1}+m_{2}+m_{3}+m_{4}+m_{5} \leq N}
b^{(j=1...4)}_{m_{1},m_{2},m_{3},m_{4},m_{5}} x_{1}^{m_{1}} x_{2}^{m_{2}} x_{3}^{m_{3}} x_{4}^{m_{4}} y_{5}^{m_{5}},
\end{equation}
\end{center}

\begin{center}
\begin{equation}\label{e_invert2}
x_{5} = \displaystyle\sum_{1 \leq m_{1}+m_{2}+m_{3}+m_{4}+m_{5} \leq N} b^{(5)}_{m_{1},m_{2},m_{3},m_{4},m_{5}}
x_{1}^{m_{1}} x_{2}^{m_{2}} x_{3}^{m_{3}} x_{4}^{m_{4}} y_{5}^{m_{5}},
\end{equation}
\end{center}

which relate the new input coordinates ($x_{i}$, $y_{5}$) with $i = 1...4$ to the new output coordinates ($y_{j}$,
$x_{5}$) with $j = 1...4$ are of interest. In the first step, Eq. (\ref{e_higher}) for $j = 5$ is entered into Eq.
(\ref{e_invert2}) and, order by order, a system of $M$ linear equations is constructed from which the coefficients
$b^{(5)}$ are determined by simple matrix inversion. In the second step, Eq. (\ref{e_invert2}) is entered into Eq.
(\ref{e_higher}) for $j = 1...4$ and by a comparison with Eq. (\ref{e_invert1}) in a similar fashion as in the first
step, the coefficients $b^{(1...4)}$ can be deduced.

Table \ref{t_order} shows the dimension $M$ of the system of linear equations which has to be solved when exchanging
one coordinate. The dimension $M$ is given as function of the order of the ion-optical matrix (assuming five input and
output variables in the ion-optical matrix).

\begin{table}[p]
 \centering
 \caption{Dimension $M$ of the system of linear equations which has to be solved when exchanging one coordinate. $M$ is a
 function of $N$, which is the order of the ion-optical matrix. Five input and output variables are assumed.}
 \begin{tabular}{|c|c|c|c|c|c|}
 \hline
 order $N$ & 1 & 2 & 3 & 4 & 5\\\hline
 dimension $M$ & 5 $\times$ 5 & 15 $\times$ 15 & 35 $\times$ 35 & 70 $\times$ 70 & 126
 $\times$ 126\\
 \hline
 \end{tabular}\label{t_order}
\end{table}

A program has been developed using the algorithm for exchanging coordinates in ion-optical matrices of five input and
output coordinates up to third order as described above. Using this program, forward ion-optical matrices were
transformed into inverse ion-optical matrices. Since the $x^{(T)}$ position is available as an input, $x^{(D)}$ and
$\theta_{x}^{(D)}$ are exchanged with $\theta_{x}^{(T)}$ and $\delta^{(T)}$. In addition, $y^{(D)}$ and
$\theta_{y}^{(D)}$ are exchanged with $y^{(T)}$ and $\theta_{y}^{(T)}$. After the exchanges, the partial inverse
ion-optical matrix is obtained as

\begin{equation}\label{e_final}
\mat{\theta_{x}^{(T)}\\ y^{(T)}\\ \theta_{y}^{(T)}\\
\delta^{(T)}\\ \Delta L^{(D)}}_{\mathrm{output}} = M
\mat{x^{(D)}\\ \theta_{x}^{(D)}\\ y^{(D)}\\ \theta_{y}^{(D)}\\
x^{(T)}}_{\mathrm{input}}.
\end{equation}

The final result can be checked against the fully inverted ion-optical matrix provided by \textsc{cosy}, because
reducing the matrix by the fifth row and the fifth column should produce the same result as \textsc{cosy} coefficient
by coefficient. It is important to note that the first exchange of two coordinates in the dispersive as well as the
non-dispersive plane should be done for a pair of coordinates which shares a large (ideally greater than one)
first-order matrix element (the \emph{pivot}).

\section{Theoretical Comparisons}
In the partial-inversion process, forward ion-optical matrices with terms up to third order are transformed into
inverse ion-optical matrices. The inverse ion-optical matrices also have terms up to third order. However, a completely
inverse ion-optical matrix would have fourth and higher order terms. Before comparing the standard \textsc{cosy} fully
inverted ion-optical matrix to the one obtained by the partial-inverse method, the impact of not including these higher
order terms is quantified.

\begin{figure}
  \begin{center}
  \includegraphics[width=\textwidth]{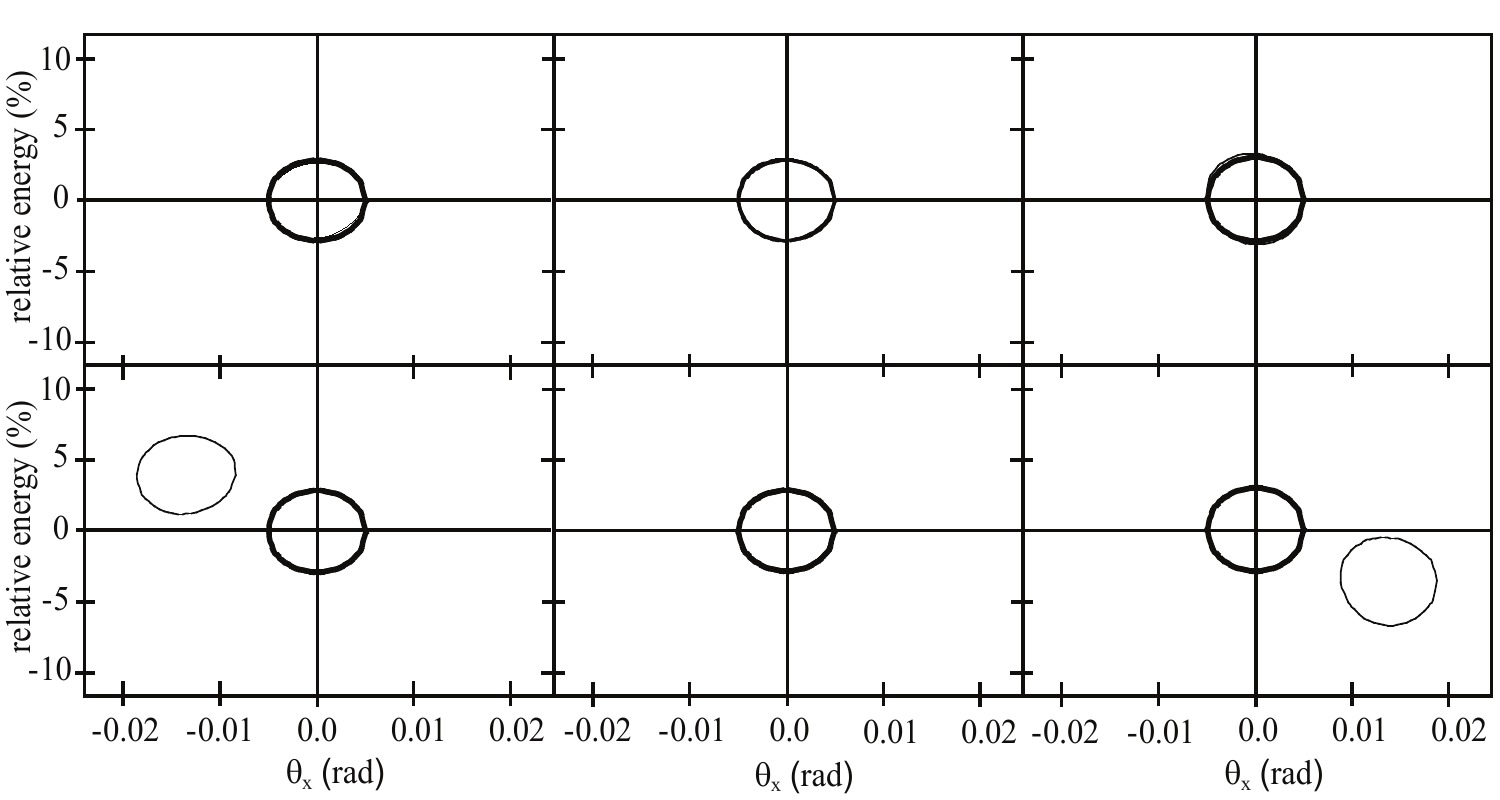}\\
  \end{center}
  \caption{Contours of relative energy versus target angle. The input contours (thick lines) are compared to output contours
  (thin lines) for the partial-inverse method (top) and the \textsc{cosy} method (bottom). The three columns are for $x^{(T)}$ positions
  (in cm) of $-1$ (left), $0$ (center), and $1$ (right).}\label{f_contour}
\end{figure}

Figure \ref{f_contour} shows comparisons of input (thick lines) to output (thin lines) contours of the relative energy
versus target angle using the partial-inverse method (top) and the \textsc{cosy} method (bottom) for three different
$x^{(T)}$ positions. The output contours are produced by propagating the input contour through a forward transformation
and then an inverse transformation using either the partial-inverse or the \textsc{cosy} fully inverted ion-optical
matrix. The input contours correspond to a spread in $\delta^{(T)}$ ($\pm$3\%) and angle $\theta_{x}^{(T)}$
($\pm$5~mrad) consistent with the characteristics of rare-isotope beams produced in fast fragmentation reactions. The
top row shows that the differences due to the lack of fourth and higher order terms for the partial-inverse method are
very small. However, with the exception of the contour pair at 0~cm in $x^{(T)}$, the comparisons show large deviations
for the \textsc{cosy} method. It should be pointed out that the difference between input to output contours increases
with the size of the contour regardless of the method chosen. In general, any transformation not performed up to all
orders will always show some deviation.

\section{Comparison with Data}
The two different reconstruction methods were compared to actual data from an experiment performed at the National
Superconducting Cyclotron Laboratory (NSCL) at Michigan State University. A primary beam of 140 MeV/u $^{40}$Ar
impinged on a 893~mg/cm$^{2}$ Be production target. The desired secondary beam of $^{26}$Ne at 86 MeV/u was purified
and delivered to the experimental setup using the A1900 fragment separator \cite{a1900_ref}. The secondary beam
interacted with the reaction target at the experimental setup resulting in charged fragments and neutrons. The charged
particles were bent out of the way using a large gap sweeper magnet built at the National High Magnetic Field
Laboratory (NHMFL) at Florida State University (FSU) \cite{sweep_refs}. The neutrons continued forward at zero degrees
and were detected with the Modular Neutron Array (MoNA) \cite{mona_ref1,mona_ref2}. The experiment was a 2p-n stripping
reaction from a $^{26}$Ne secondary beam to neutron unbound states in $^{23}$O as described in \cite{sch_ref,nate_ref}.

\begin{figure}
  \begin{center}
  \includegraphics[width=0.8\textwidth]{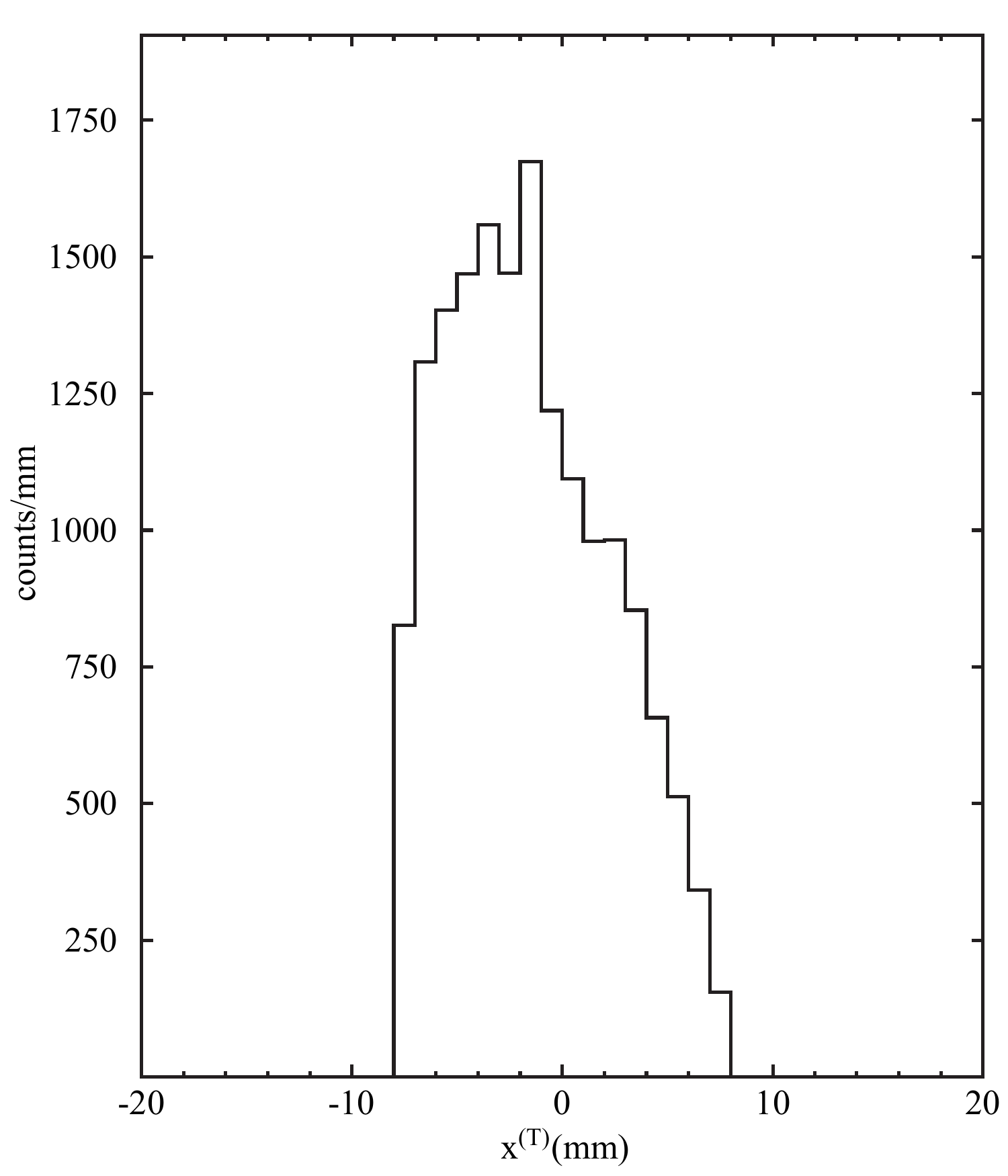}\\
  \end{center}
  \caption{Dispersive position distribution $x^{(T)}$ of beam particles at the
  reaction target.}\label{x_target}
\end{figure}

The beam position at the target was determined by propagating a measured trajectory through a quadrupole triplet
ion-optical matrix. The measured trajectory was determined by two position detectors before the quadrupole triplet. The
beam position and size in the dispersive direction is shown in Fig. \ref{x_target}. The beam was offset from the center
by approximately 3~mm and had a width of 20~mm.

The relative energy $\delta$ and the target angle $\theta_{x}^{(T)}$ provide the best comparison between both
reconstruction methods because they have matrix elements in first order that depend on $x^{(T)}$. Data were taken
without a target so that the reconstructed energy could be compared directly with the energy of the incoming $^{26}$Ne
beam which was determined from a time-of-flight measurement between two scintillators located in the beamline separated
by 35.7~m.

\begin{figure}
  \begin{center}
  \includegraphics[width=\textwidth]{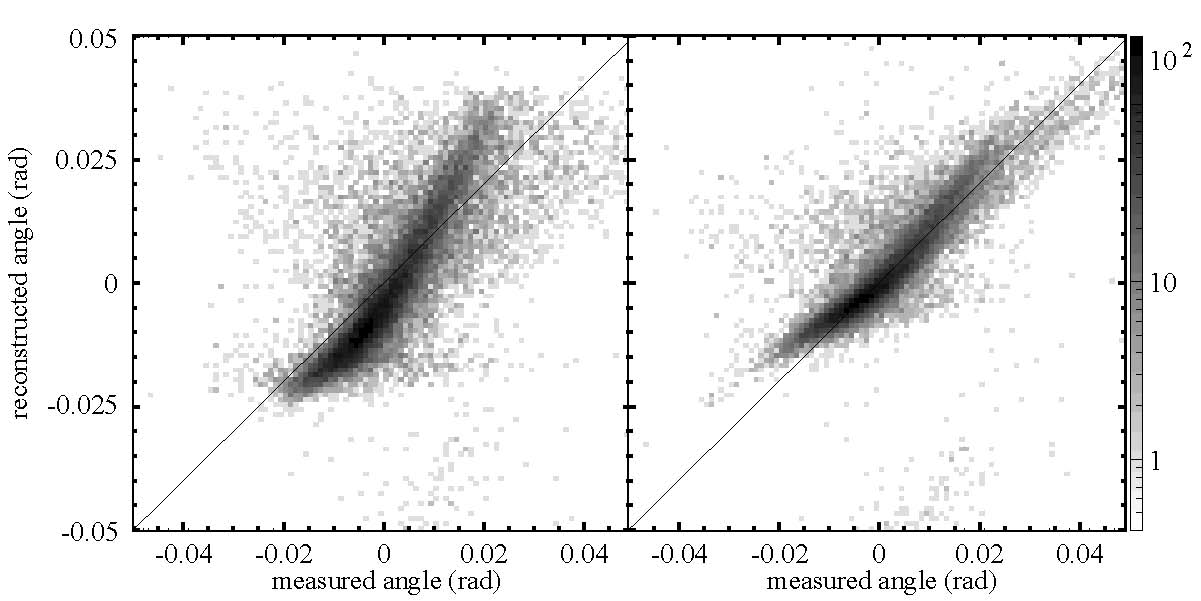}\\
  \includegraphics[width=\textwidth]{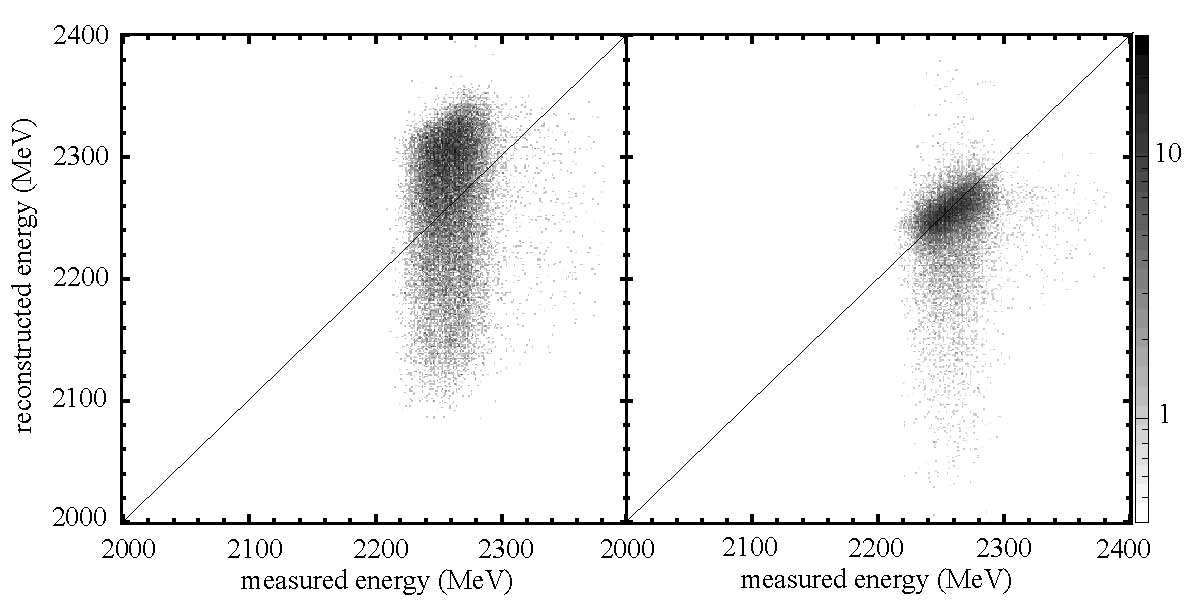}\\
  \end{center}
  \caption{Reconstructed versus measured target angle $\theta_{x}^{(T)}$ (top) and reconstructed versus measured energy (bottom)
  distributions for the \textsc{cosy} method (left) and the partial-inverse method (right). The diagonal lines represent a one-to-one
  correspondence.}
  \label{f_thetax}
\end{figure}

Figure \ref{f_thetax} shows a comparison of the reconstructed versus measured target angle $\theta_{x}^{(T)}$ (top) and
reconstructed versus measured energy (bottom) distributions for the \textsc{cosy} method (left) and the partial-inverse
method. The distributions from the \textsc{cosy} methods show significantly larger deviations from the ideal one-to-one
correspondence indicated by the solid diagonal line than the partial-inverse method.

\begin{figure}
  \begin{center}
  \includegraphics[width=0.7\textwidth]{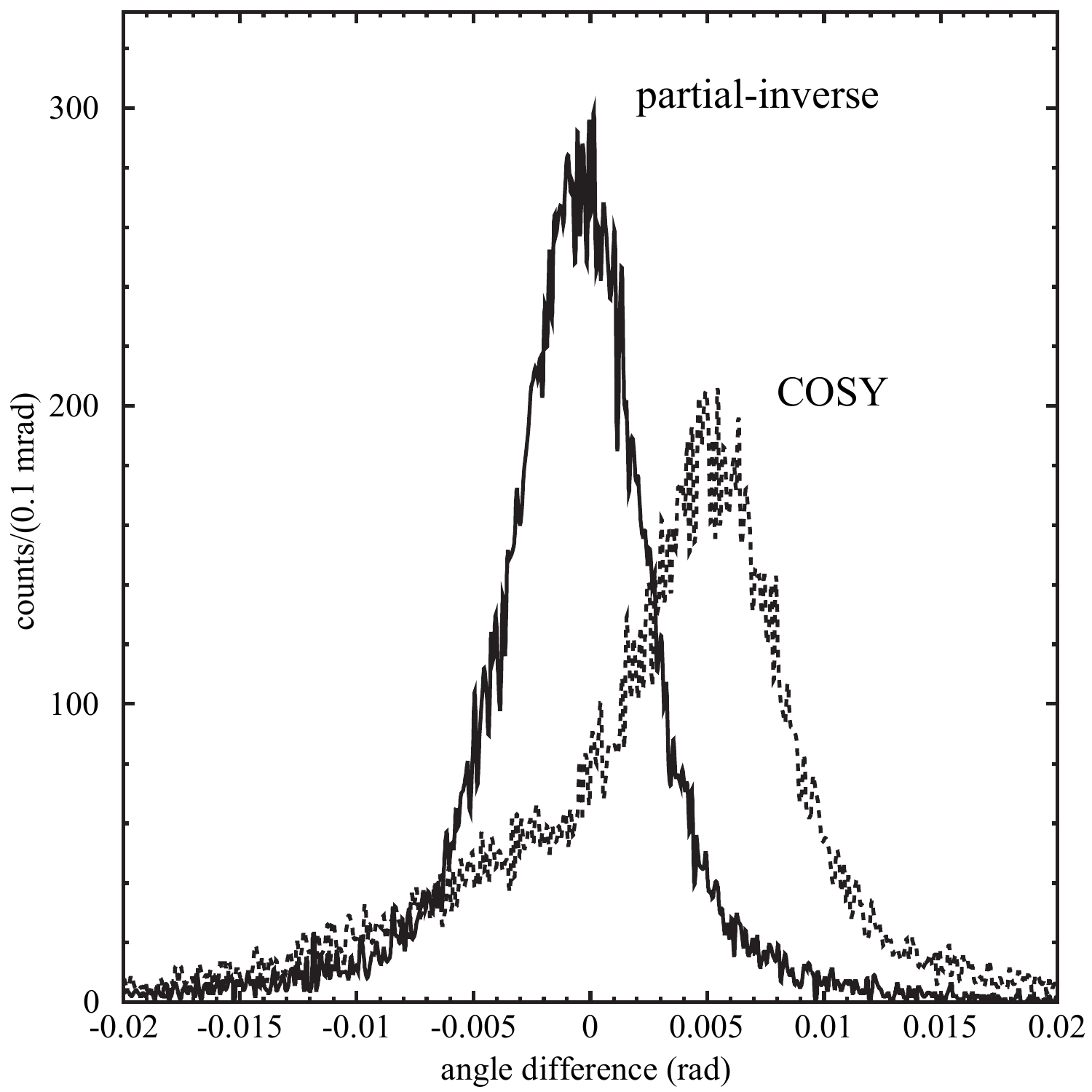}\\
  \end{center}
  \caption{Difference between measured and reconstructed $\theta_{x}^{(T)}$ at the reaction target
  for the partial-inverse (solid) and \textsc{cosy} (dashed) methods.}
  \label{f_thetax_diff}
\end{figure}

\begin{figure}
  \begin{center}
  \includegraphics[width=0.7\textwidth]{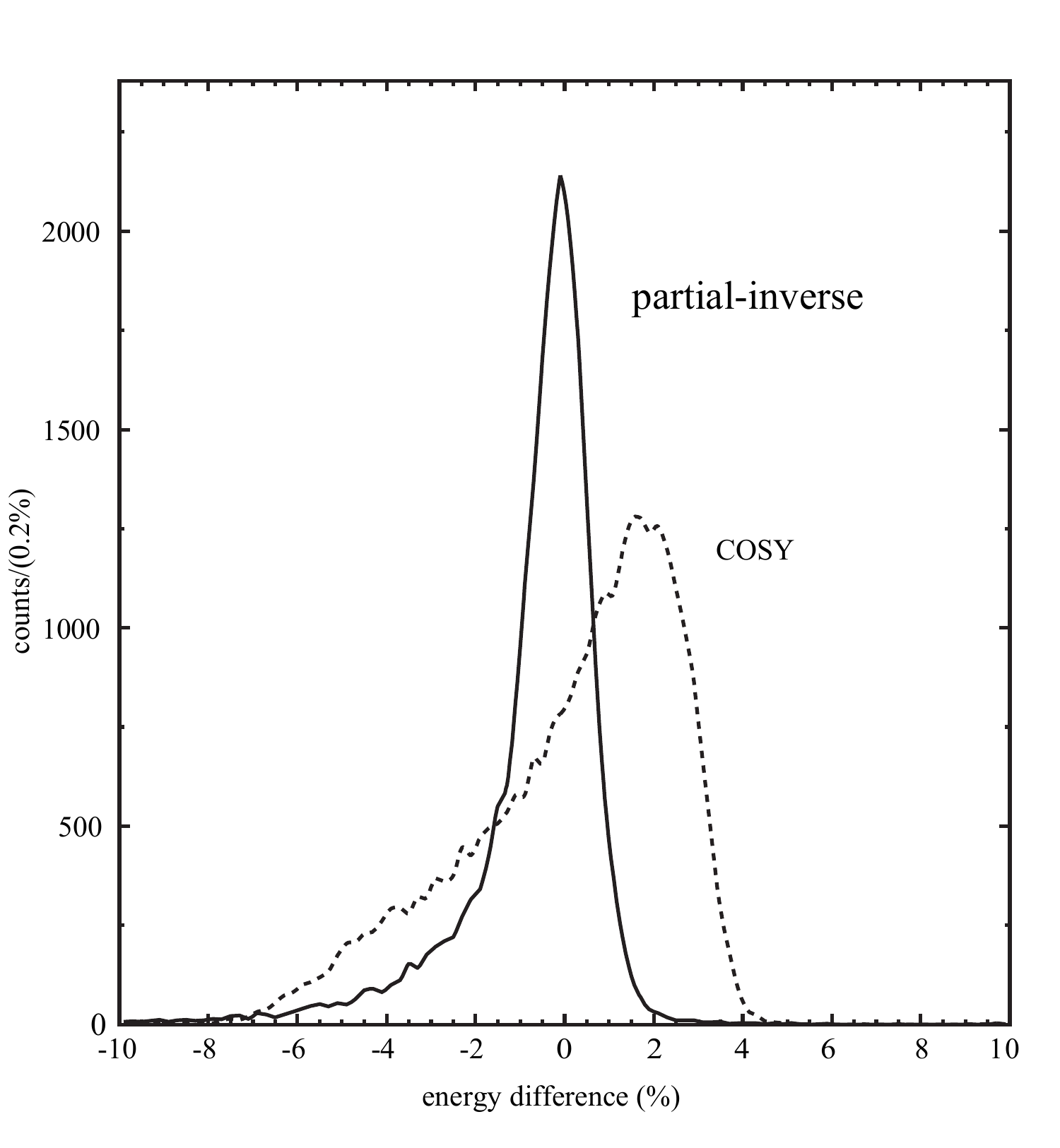}\\
  \end{center}
  \caption{Difference between the measured and reconstructed beam energy at the reaction target position
  for the partial-inverse (solid) and \textsc{cosy} (dashed) methods.}
  \label{f_e_diff}
\end{figure}

A more quantitative comparison is shown in Figs. \ref{f_thetax_diff} and \ref{f_e_diff} where the differences between
the measured and the reconstructed target angle and energy, respectively, are plotted. The centroid of
$\theta_{x}^{(T)}$ distribution is not reproduced by the fully inverse \textsc{cosy} matrix and differs by
$\sim$6~mrad. The improvement for the partial-inverse method is obvious as the centroid of the distribution agrees with
the measured data. In addition, the result of the partial-inverse method yields a narrower width for the difference
distribution of $\theta_{x}^{(T)}$.

Even more important is the improvement for the reconstruction of the energy. Figure \ref{f_e_diff} shows the deviation
of the reconstructed energy from the measured energy in \%. On average, a $\sim$2\% deviation in beam energy is
observed for the \textsc{cosy} method, while the partial-inverse method reconstructs the energy correctly. Also, the
FWHM of the distribution using the partial-inverse method is only $\sim$1.5\% as compared to $\sim$4\% using the
\textsc{cosy} method.

\section{Conclusion}
The partial-inverse method significantly improves the accuracy of the reconstruction of four-momenta for experiments
using secondary beams which have typically large beam spots. The assumption of an infinitely small beam spot size used
in the fully inverse \textsc{cosy} method is insufficient to extract spectroscopy information from the data. The
partial-inverse method utilizes the information of the finite beam spot size for the correct reconstruction. Thus, in
order to extract the optimum resolution by using this method it is necessary to measure the beam position with
appropriate tracking detectors.

\section{Acknowledgments}
The authors wish to thank the National High Magnetic Field Laboratory and the National Superconducting Cyclotron
Laboratory for the work in construction, testing, and installation of the Sweeper Magnet. This work is supported by the
National Science Foundation grant PHY-06-06007.

\end{document}